# Unsupervised Threat Hunting using Continuous Bag-of-Terms-and-Time (CBoTT)

*Completed Research Paper*

**Varol Kayhan[1], Shivendu Shivendu[2],**
**Rouzbeh Behnia[3], Clinton Daniel[4], Manish Agrawal[5]**
University of South Florida
4202 E. Fowler Ave, Tampa, FL, 33620, USA
[1] vkayhan@usf.edu
[2] shivendu@usf.edu
[3] behnia@usf.edu
[4] cedanie2@usf.edu
[5] magrawal@usf.edu

## Abstract

*Threat hunting is sifting through system logs to detect malicious activities that might have bypassed existing security measures. It can be performed in several ways, one of which is based on detecting anomalies. We propose an unsupervised framework, called continuous bag-of-terms-and-time (CBoTT), and publish its application programming interface (API) to help researchers and cybersecurity analysts perform anomaly-based threat hunting among SIEM logs geared toward process auditing on endpoint devices. Analyses show that our framework consistently outperforms benchmark approaches. When logs are sorted by likelihood of being an anomaly (from most likely to least), our approach identifies anomalies at higher percentiles (between 1.82-6.46) while benchmark approaches identify the same anomalies at lower percentiles (between 3.25-80.92). This framework can be used by other researchers to conduct benchmark analyses and cybersecurity analysts to find anomalies in SIEM logs.*

**Keywords**: threat hunting, anomaly detection, continuous bag-of-words, API

## Introduction

According to the Identity Theft Resource Center's Annual Data Breach Report, organizations reported 1,862 data breaches in 2021, a 68% increase from the previous year (Identity Theft Resource Center, 2022). In a typical data breach, a malicious actor creates or exploits a vulnerability, enters an organization's network, starts looking for data, and transfers the data out of the network inconspicuously (Nelson & Simek, 2019). A recent report from IBM estimated that around 83% of the surveyed organizations reported more than one data breach (IBM.com, 2022b). This is due in part to the increasing involvement of human adversaries in cybersecurity threats. These adversaries leverage normal technology behaviors to evade detection by individual cybersecurity solutions such as endpoint protection or firewalls. Unlike earlier threats, such as malware, which were primarily based on malicious code and exploited known software vulnerabilities, human attackers today are increasingly exploiting regular software and utilities to accomplish their objectives.

To detect adversarial activities, one approach is to collect logs using security information event management (SIEM) tools. These tools aggregate logs collected by endpoint systems and use signatures of known-bad events to identify and detect malware and common exploits. However, they provide limited defense against sophisticated attackers who use novel techniques such as zero-day exploits and stealthy procedures (Bowman et al., 2020). Therefore, while SIEM tools are useful, they have limitations and require extensive manual effort to integrate logs from different security tools and vendors. Additionally, the resulting delay and variability in processing SIEM logs can leave most sophisticated attacks and intrusions undetected until it is too late, which can damage organizations' data and reputation.

One proactive approach to preventing cybersecurity breaches and cyberattacks is *threat hunting*, which involves analyzing system logs to detect any malicious activity that may have evaded existing security measures (rThreat, 2021; SQRRL, 2021; Trellix.com). Threat hunting aims to identify threats that remain undetected in an organization despite various cybersecurity controls. Such threats could be in the form of persistent executables that remain on devices and employ or elevate user permissions to steal information. In this regard, threat hunting is a specialized case of anomaly detection and often requires a human-centric approach.

Threat hunters rely on their understanding of an organization's typical technology-use patterns to identify indicators of compromise (IoCs) by looking for harmful deviations from these patterns. These IoCs may include information such as IP addresses, port numbers, startup services, DNS queries, running processes, opened files, accessed registry keys, and executed commands. However, as the number of alerts increases, analyzing them becomes both ineffective and costly, as it requires significant human effort and time. As a result, there is a growing interest in using artificial intelligence (AI) and machine learning (ML) methods for threat hunting. According to a recent survey by SANS, the use of AI/ML tools for threat hunting has increased from 15% in 2020 to approximately 45% in 2022 (Fuchs & Lemon, 2022).

Threat hunting is usually performed in logs collated by security information and event management (SIEM) tools. These tools acts as a centralized repository for logs gathered from sources such as Endpoint Detection and Response (EDR) tools that are installed on endpoint devices (fireeye.com; imperva.com). EDR tools can be set up to perform various types of audits on endpoints such as process auditing, which captures every process executed by either users or applications on that endpoint. Processes can take many forms, including specific commands submitted by users through a terminal screen, as well as instances of applications or programs executed by users or the operating system.

For instance, if a user executes the command "`C:\Windows\System32\cmd.exe ipconfig -all`" to retrieve the current network configuration of an endpoint device, an EDR tool could capture this command and flag it as suspicious based on predefined rules. Similarly, if a user downloads and opens a Microsoft Excel file, called "myfile.xlsm," that has a macro in it, an EDR tool could consider this suspicious and capture the command "`'C:\Program Files\Microsoft Office\Root\Office16\Excel.exe' 'C:\Users\Downloads\myfile.xlsm'`." In both cases, the EDR tools may forward these processes to a SIEM tool for logging. As seen in both cases, a process is a text-based command that performs a specific task or starts an application or program. Despite their usefulness, EDR tools pose two types of challenges for threat hunting. First, they perform auditing based on predefined rules. As a result, they may generate too many false positives (Giannetti, 2022; Ulevitch, 2017), making it difficult for analysts to sift through large volumes of data. Second, because processes are textual commands, analysts must examine them one at a time to determine whether they pose any risk. Therefore, it is crucial to develop machine learning based tools and techniques that can assist analysts in quickly and automatically analyzing text-based commands.

The goal of this paper is to address these two challenges by leveraging recent advancements in machine learning and, specifically, natural language processing. We propose a new framework called *continuous bag-of-terms-and-time* (CBoTT) to enable cybersecurity analysts to process large volumes of logs containing text-based process audits and determine if there are any processes that pose security risks. Our framework is an extension of the popular continuous bag-of-words approach (Mikolov et al., 2013) and enables us to identify the processes that should be investigated with respect to not just what they do, but also when they are executed.

## Motivation

According to IBM's 2022 data breach report, the average cost of a data breach in the US is $9.44 million (IBM.com, 2022a). The same report indicates that it takes about nine months, on average, to discover and identify a breach, but if organizations lack the necessary tools or methods to conduct investigations, it can take much longer to identify data breaches. For example, the Marriott data breach that put 500 million

people at risk was discovered four years after the breach happened (Kerner, 2018). Organizations with the right tools and techniques not only reduce the time it takes to discover and respond to data breaches, but also save an average of $3 million compared to those without them (IBM.com, 2022a). Therefore, many information systems scholars call for the development of tools, techniques, and frameworks to fight malicious actors (Baskerville, Spagnoletti, & Kim, 2014; Greengard, 2016; Kappelman et al., 2019). In particular, information systems researchers are increasingly relying on artificial intelligence and machine learning to combat threats (e.g., Abbasi et al., 2021; Ebrahimi et al., 2022; Ebrahimi, Nunamaker Jr, & Chen, 2020; Li, Chen, & Nunamaker Jr, 2016; Samtani et al., 2017).

Threat hunting is an approach to identify breaches or cyberattacks by analyzing system logs to detect malicious activities that may have evaded existing security measures (rThreat, 2021; SQRRL, 2021; Trellix.com). However, this approach presents a challenge due to the high volume of logs generated by each endpoint device, especially for large organizations where there could be many endpoints. (Kent & Souppaya, 2006). For instance, an average enterprise can generate up to 4GB of log data a day, making the volume of data overwhelming (Progress Software). If analysts conduct threat hunting for one month's worth of data, they must investigate 120GB of data, which can be very difficult without using any automated tools. Moreover, most of this data comprises normal events, leading to a high number of false positives. For instance, in a recent study, where security operations center (SOC) analysts are interviewed, one analyst reports that 99% of the data they analyze are false positives (Alahmadi, Axon, & Martinovic, 2022). To address this challenge, there is a need for new tools, techniques, or frameworks that can efficiently examine large volumes of log data and identify suspicious logs that may pose a threat to an organization. In this study, we propose a new framework that draws upon recent advancements in natural language processing and machine learning to achieve this goal.

## *Novelty*

Our framework is different from the current state-of-the-art in several ways. First, our approach for threat hunting is unsupervised and does not require labeled data. This addresses a critical limitation since many of the extant studies (e.g., Goldstein et al., 2013; Guo, Yuan, & Wu, 2021; Haque, DeLucia, & Baseman, 2017; Li et al., 2018; Wittkopp et al., 2021) require the domain specific data to be labeled by cybersecurity analysts for training. This requires the analysts to sift through millions of logs and build a training data set that distinguishes between normal and unusual behavior. Given the volume of logs generated by SIEM tools, this is impractical, slow, and costly. More importantly, even if such a data set is created, it might change over time even in the same organization because of the dynamic nature of an organization's business, processes, and technologies.

Second, our framework calculates the probability that each observation requires investigation by a cybersecurity analyst. This allows cybersecurity analysts to sort the results of the framework by probability, and thus, start investigating those that pose the greatest risk. This not only helps analysts attend to those observations that might need immediate attention, but also enables them to determine the depth of their investigation. For example, in one context, a cybersecurity analyst might choose to examine the top one percent of the sorted results, while in another, they might choose to investigate the top five percent. To the contrary, most, if not all, existing approaches provide a binary outcome to only suggest whether an observation requires investigation. This can make analysts more likely to miss the false negatives because they might not know where to look for these observations among results.

Third, our framework can be used at the endpoint level, allowing analysts to account for differences between business functions. For instance, it can analyze the processes of an IT endpoint separately from those of a business endpoint and identify the processes that require investigation based on the endpoint-level differences. This sets our approach apart from those proposed in earlier studies that analyze data sets with no regard to differences among endpoints.

In the pages that follow, we first discuss the background and prior work in this area. Then, we provide an in-depth explanation of the CBoTT framework. After that, we discuss the framework's performance using an example data set and compare it to various benchmark models. Finally, we present our concluding remarks.

# Background and Prior Work

## *Threat hunting*

Threat hunting can be defined as *searching through system logs to identify malicious activities that evade existing security solutions* (Kerwin, 2021; rThreat, 2021; SQRRL, 2021). It is performed in at least three ways: 1) intel-based, 2) hypothesis-based, and 3) anomaly-based (IBM.com). In this study, we adopt the anomaly-based approach to perform threat hunting in SIEM logs collected from endpoint devices. This approach is a form of anomaly detection, which involves identifying observations that deviate from other observations in a data set (Hawkins, 1980). It is important to note that anomalies are usually defined by frequency of occurrence. While less frequently occurring observations are considered anomalies, one can also consider more frequently occurring observations as anomalies. In this study, we focus on less frequently occurring ones because these are usually more difficult to identify in logs. It is also worth mentioning that an anomaly might not necessarily be malicious or pose a threat to an organization, but it indicates that a command is different from other commands has been executed on an endpoint device. As a result, we consider anomalies as observations that pose risks to an organization, and therefore, present them to cybersecurity analysts for further investigation.

When EDR tools conduct process auditing, traditional anomaly detection techniques cannot be used to analyze logs due to the unstructured nature of the data. Further, simple enumeration-based techniques, such as sorting the data in alphabetical order and counting the number of times each command occurs in the data, may not provide useful insights due to different variations of the same command. For example, consider the command "`netsh advfirewall show currentprofile`," which retrieves the status of Windows Firewall for the current profile. This same task can be performed using various other commands, such as "`C:\Windows\System32\cmd.exe netsh advfirewall show currentprofile`" or "`powershell.exe netsh advfirewall show currentprofile`." There could be many other variations of this command as well, such as "`cmd.exe netsh advfirewall show currentprofile > tmp001.txt`," which omits the full path and writes the output to a temporary file called "`tmp001.txt`." Despite their different syntax, these commands all perform the same task. Therefore, sorting commands alphabetically and counting their frequency is not useful, and neither is using regular expressions. While regular expressions can be used to standardize commands and create templates, this approach requires knowledge of what is considered normal. To overcome these challenges and perform anomaly detection in text-based data, researchers use machine learning and artificial intelligence algorithms. A review of sample work conducted in this area is presented in Table 1.

As seen in Table 1, prior work conducts both unsupervised (e.g., Baseman et al., 2016; Li et al., 2018; Meng et al., 2019; Xu et al., 2009) and supervised learning (e.g., Goldstein et al., 2013; Guo, Yuan, & Wu, 2021; Haque, DeLucia, & Baseman, 2017; Li et al., 2018; Wittkopp et al., 2021) to find anomalies in text-based data. Most studies examine logs files generated by HPC technologies (Baseman et al., 2016; Guo, Yuan, & Wu, 2021; Haque, DeLucia, & Baseman, 2017; Li et al., 2018; Meng et al., 2019; Wittkopp et al., 2021). There is also work that focuses on Hadoop Distributed File System (HDFS) logs (Du et al., 2017; Guo, Yuan, & Wu, 2021; Meng et al., 2019; Xu et al., 2009) as well as SIEM logs (Goldstein et al., 2013). Most studies extract patterns using predefined log templates generated using regular expressions (e.g., Du et al., 2017; Guo, Yuan, & Wu, 2021; Li et al., 2018; Meng et al., 2019; Wittkopp et al., 2021). These templates are used to identify anomalous sequences of events.

The types of approaches used to detect anomalies include principal component analysis (Xu et al., 2009), nearest neighbors (Goldstein et al., 2013), graph analysis (Baseman et al., 2016), Markov chain models (Haque, DeLucia, & Baseman, 2017), neural networks (Du et al., 2017; Li et al., 2018), word2vec (Meng et al., 2019), and transformers (Guo, Yuan, & Wu, 2021; Wittkopp et al., 2021). To evaluate the goodness of models, prior work mostly uses F1-score, precision, and recall (e.g., Du et al., 2017; Guo, Yuan, & Wu, 2021; Haque, DeLucia, & Baseman, 2017; Li et al., 2018; Meng et al., 2019; Wittkopp et al., 2021; Xu et al., 2009). However, there are also studies that use ranking to identify each observation's likelihood of being an anomaly (e.g., Baseman et al., 2016; Goldstein et al., 2013).

| Study | Supervised/ Unsupervised | Data | Type of anomaly detection | Approach | Evaluation |
|---|---|---|---|---|---|
| Guo et al. (2021) | Supervised | High performance computing system logs & Hadoop distributed file system logs | Collective | Transformer based | F1-score, precision, recall |
| Wittkopp et al. (2021) | Supervised | High performance computing system logs | Point | Transformer based | F1-score, precision, recall |
| Meng et al. (2019) | Supervised | High performance computing system logs & Hadoop distributed file system logs | Collective | Word2vec based | F1-score, precision, recall |
| Li et al (2018) | Unsupervised & supervised | High performance computing system logs | Collective | Neural network based | Accuracy |
| Du et al. (2017) | Supervised | Hadoop distributed file system logs | Collective | Neural network based | F1-score, precision, recall |
| Haque et al. (2017) | Supervised | High performance computing system logs | Collective | Markov chain based | Precision, recall |
| Baseman et al. (2016) | Unsupervised | High performance computing system logs | Point | Graph based | Ranking |
| Goldstein et al. (2013) | Supervised | System information and event management logs | Point | Nearest neighbor based | Ranking |
| Xu et al. (2009) | Unsupervised | Hadoop distributed file system logs | Collective | Principal component based | Precision, recall |

**Table 1. Summary of extant work on text-based anomaly detection**

To summarize, there has been significant research in identifying anomalies in text-based data, but there are also limitations to this work. One major limitation is that many studies separate anomalous observations from normal ones to train models on normal observations only (e.g., Goldstein et al., 2013; Guo, Yuan, & Wu, 2021; Haque, DeLucia, & Baseman, 2017; Li et al., 2018; Wittkopp et al., 2021). Despite claiming to use unsupervised learning, this approach still requires researchers or cybersecurity analysts to have prior knowledge of what is considered normal. With the large volume of logs generated by cybersecurity tools, it is challenging to distinguish between normal and anomalous observations for model training. Moreover, this leads to a continuous need to train new models as the definition of normal changes over time, or else risk creating too many false positives.

Another limitation of prior work is that it mostly focuses on the byproducts of processes rather than the actual processes themselves for anomaly detection. For instance, logs generated by HPC and HDFS are the outputs of processes executed by users. Therefore, previous studies have primarily examined the outputs of processes as captured in logs, rather than analyzing the commands that invoke these processes. In the following section, we propose our framework to address these challenges.

### Anomaly Detection using Text Mining

One of the foundational strategies for conducting anomaly detection on text-based data entails processing the data using techniques rooted in text mining (see Salton & McGill, 1983) and applying outlier detection algorithms on the processed data set. The data processing involves parsing the text using a tokenizer, followed by creating a document-term matrix from the parsed terms. In this matrix, each observation is represented as a set of vectors. Vector weights can be binary, to indicate the presence or absence of terms in an observation, or term frequency-inverse document frequency (TF-IDF) values, to capture the relative significance of a term in one observation compared to its prevalence across the entire data set. However, it is important to note that this document-term matrix often exhibits high dimensionality, which can pose challenges to machine learning algorithms. Therefore, a recommended approach is to truncate this matrix to retain only the top $N$ most frequently occurring terms in the data set (see Lam & Lee, 1999). This enables algorithms to adequately distinguish between observations without being burdened by the dimensionality of the full matrix (see Dumais et al., 1988; Xiao & Cho, 2016). This truncated matrix can be sent to outlier detection algorithms, such as one-class support vector machine, or density-based special clustering of applications with noise (DBSCAN), to identify anomalies (e.g., Antonelli et al., 2013; Najafabadi et al., 2017).

### Continuous Bag-of-words

Continuous bag-of-words (CBOW) was developed by Mikolov et al. (2013) as a method to accurately learn word vectors from massive data sets that contain billions of words. Word vectors represent words in a high-dimensional space as numerical values. For example, in a 300-dimensional vector space, each word is represented by a set of 300 numbers. The objective is to train a neural network to accurately learn these values for every word in the data set so that semantic similarities between words can be captured. This is important for natural language processing because it can lead to versatile applications such as being able to perform arithmetic operations on word vectors. For example, subtracting the vector of the word "big" from the vector of "biggest" and then adding the result to the vector of "small" generates the vector of the word "smallest" (Mikolov et al., 2013).

To achieve this, a neural network must be trained on a data set by organizing words into sets of inputs and outputs. This data set is generated by CBOW using sliding a window over a sentence that is parsed into words. In each window, the middle word is designated as the output, while a specific number of preceding and succeeding words (near the middle word) is designated as inputs. The neural network's objective is to learn the word vectors so that it can accurately classify the middle words based on the nearby preceding and succeeding terms. We adopt this CBOW approach in our framework, which is described in detail in the subsequent section.

## CBoTT: Continuous Bags-of-Terms-and-Time

The CBoTT framework extends a well-known language processing technique called continuous bag-of-words (see Mikolov et al., 2013). We believe this approach is superior to attention-based models, such as transformers, for at least three reasons. First, pre-trained transformers are designed for natural languages and may not be effective for commands that use different vocabulary or grammatical rules. When faced with unfamiliar terms or file paths that commonly occur in commands but not in natural language, tokenizers used by transformers, such as WordPiece (Schuster & Nakajima, 2012), might incorrectly break down these terms or paths to find matches in the pre-trained vocabulary, leading to misinterpretations. Additionally, the grammatical rules used in natural language might cause pre-trained transformers to struggle with interpreting the specific order in which terms are used in commands. As a result, pre-trained transformers might incorrectly identify commonly occurring commands as anomalies, leading to an excessive number of false positives in cybersecurity applications. This limitation has already been documented in the context of programming. For example, GPT-3, a large language model trained on natural language, cannot solve any of the programming problems specifically designed to test its code-writing capabilities due to the disparities between programming languages and natural language with respect to vocabulary and rules (see Chen et al., 2021).

Second, training a transformer from scratch to detect anomalies requires labeled data (e.g., Guo, Yuan, & Wu, 2021; Wittkopp et al., 2021), which is time-consuming and impractical for cybersecurity analysts. This is because analysts must either obtain these data sets from external sources, which could impact the applicability of models to different contexts or dedicate considerable effort and time to create one. Additionally, even if a data set is successfully created, analysts must continually update it to account for new norms, commonly referred to as *concept* or *data drift* in the context of machine learning (Widmer & Kubat, 1996). Otherwise, models trained on outdated data might generate more anomalies, exacerbating the false-positive issue in cybersecurity.

Third, while transformers can learn the structure of text, they are limited in their ability to learn other metadata about the text. For instance, a transformer can learn the structure of a command, but it may not learn when the command is executed. This is especially critical in situations where the command is executed outside of regular working hours. Malicious actors who infiltrate a network typically take their time in exfiltrating data from the network over a period of several weeks or months to avoid detection (Nelson & Simek, 2019). These activities may occur outside of regular business hours and even during weekends. Therefore, the execution times of tasks can provide valuable insights into unusual behaviors. Thus, we believe our approach is better suited for the task at hand. In the next section, we provide an overview of our approach.

## *Overview*

A summary of our framework is presented in Figure 1. As seen in the figure, a SIEM tool collects process auditing logs from each endpoint device, referred to as a *host* hereafter, and aggregates them in a single repository. We extract the command that initiates each process from these logs. Then, we group the processes of each host using a sliding window approach to identify the *tasks* executed at that host. Using the continuous bag-of-words approach, we create a separate training dataset for each host and train a word-embedding-based neural network model. These models learn the representations of tasks as well as their execution times at each host. Next, we use the trained models and the same data sets to make inferences. Rather than making predictions, we retrieve each model's output arrays and identify actual terms' and their actual execution times' probabilities. We use these probabilities to calculate each task's and its execution time's probability of occurrence at its host. Low probabilities indicate that a task is an anomaly and less likely to be executed at that time at its host.

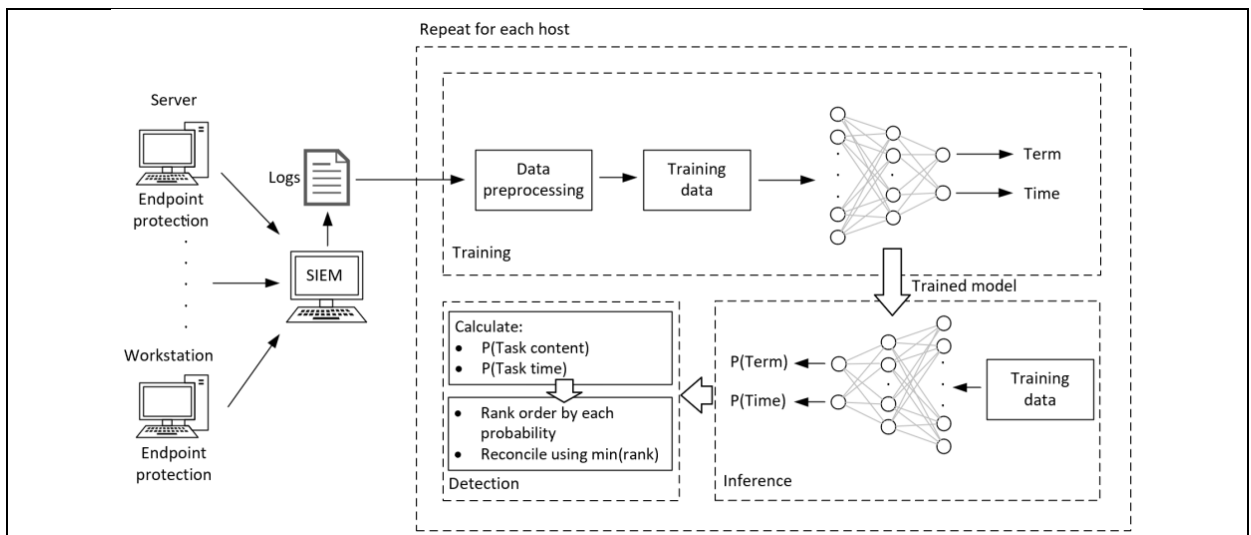

**Figure 1. Overview of the CBoTT Framework**

### Data Preprocessing

The first step of our framework is data preprocessing. To identify the terms of a command, we use a tokenizer that converts all characters to lowercase and replaces separators and punctuation characters (such as forward and backward slashes, single and double quotes, and backquote) with the space character. Using the space character as the delimiter, we identify each term of every command. We define a parameter $\gamma$ to limit the number of terms considered for each command. If a command has more terms than $\gamma$, those terms are dropped from analysis. For example, consider the command "`C:\Windows\System32\cmd.exe ipconfig /all`." Assuming that $\gamma > 6$, the tokenizer identifies the following terms from this command: "`c`" "`windows`" "`system32`" "`cmd.exe`" "`ipconfig`" "`all`."

After tokenization, we identify the *tasks* performed on each host by grouping the individual commands executed at that host. A task can consist of many commands since individual commands often do not make sense without the context of others. We create tasks rather than examining each command separately because commands executed in a window of time may be linked to each other and provide insights into the nature of tasks executed at each host. For example, consider a malicious actor who compromises a host and wants to move laterally within the network to access sensitive and valuable information. The actor would need to execute multiple commands to identify the characteristics of the current host and other hosts in the network. By grouping these commands together, it is possible to gain a better understanding of the actor's intent than by examining them individually. This approach has been used extensively in other domains to understand users' intentions. For example, consider web analytics, where user engagement with websites is analyzed using sessions. In this case, each session is a group of interactions because individual interactions alone do not provide insight into how users engage with websites (e.g., Liu, Kumar, & Mookerjee, 2020; Sun et al., 2022). Similarly, grouping commands into tasks enables us to shed light on the activities performed on each host and identify unusual activities.

Therefore, we first identify each host $H_i$ where $i \in [1, m]$. Each host $H_i$ has tasks $Y_{i,j}$, where $j \in [1, n]$ denotes the task number, and $i$ refers to the host $H_i$ the task belongs to. We define each task as a series of commands $c_{j,k}^{H_i}$ where $k \in [1, l]$. Task boundaries are determined by an amount of inactivity $\delta$ between two consecutive commands $c_{j,k}^{H_i}$ and $c_{j,k+1}^{H_i}$. Evidently, a smaller value of $\delta$ results in identifying tasks with smaller number of commands. Further, we mark the transition from one command to the next (i.e., `<nextcommand>`) as well as the beginning and end of each task (i.e., `<begintask>` and `<endtask>` respectively).

### Training Set

After data preprocessing, we develop a separate training set for each host as follows. Using the continuous bag-of-words approach, we use a sliding window approach (of size $w$ where $w \in [3,5,7,\dots]$ and stride=1) on the terms of each task from beginning to end. In each window, the middle term, $\tau_w$, is identified as the first target variable while the surrounding terms make up the input variables. In the same window, we further identify a second target variable, $\tau_t$, to capture the time (in hours) at which this term is executed. Because each task consists of multiple processes—which might have different execution times—the API calculates the mean execution time for each task and uses its hour as the value of this target variable.

| Inputs | | | | Target1 ($\tau_w$) | Target2 ($\tau_t$) |
|---|---|---|---|---|---|
| 0 | 0 | c | windows | `<begintask>` | 16 |
| 0 | `<begintask>` | windows | system32 | c | 16 |
| `<begintask>` | C | system32 | cmd.exe | windows | 16 |
| c | Windows | cmd.exe | ipconfig | system32 | 16 |
| windows | system32 | ipconfig | all | cmd.exe | 16 |
| system32 | cmd.exe | all | `<endtask>` | ipconfig | 16 |
| cmd.exe | Ipconfig | `<endtask>` | 0 | all | 16 |
| ipconfig | All | 0 | 0 | `<endtask>` | 16 |

**Figure 2: Example data set created from a single-command task.**

As an example, consider a tokenized task that consists of a single command, such as "`<begintask> c windows system32 cmd.exe ipconfig all <endtask>`" that was executed at `01-15-2023 16:15:00`. If $w$=5, the inputs and target variables are presented in Figure 2.

## Model

We use the training set generated in the previous step to train a multi-target prediction model for each host. This allows the model to learn the middle terms as well as their execution times simultaneously.

This model is a four-layered neural network model as shown in Figure 3. Because the terms cannot be processed as they appear in the data, the neural network uses an embedding layer to generate word embeddings. This layer uses a transformation function $g : V_i \rightarrow \mathbb{R}^d$ (where $V$ represents the entire vocabulary observed in the data set) that converts each term to a vector of real values $\vec{v_j}$. The embedding layer has an output dimension of $d$. Therefore, $\vec{v_j}$ is a vector of size $d$. The embeddings are trainable in the neural network using back propagation.

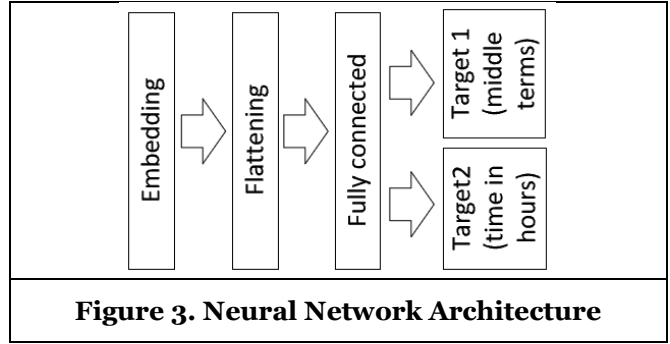

**Figure 3. Neural Network Architecture**

In the next layer, we flatten the output of the embedding layer to create the feature set $f \in \mathbb{R} = F$. This is sent to a fully connected layer using the following function, where $\sigma$ is the activation function, $W$ is the weight matrix, and $b$ is the bias matrix:

$$h = \sigma(Wf + b)$$

The final layer consists of two separate fully connected output layers. While the layer labeled as "Target1" is trained on the embedding vectors of middle terms, the layer labeled as "Target2" is trained on the execution time of terms. Both layers use softmax as the activation function:

$$Target1\ softmax \quad \sigma(\vec{w})_i = \frac{e^{w_i}}{\sum_{j=1}^{Y} e^{w_j}} \quad , \quad Target2\ softmax \quad \sigma(\hat{t})_i = \frac{e^{t_i}}{\sum_{j=1}^{Y} e^{t_j}}$$

We use categorical cross entropy as the loss function for both output layers. Therefore, the neural network is trained using the sum of two categorical cross entropy functions to account for both output layers during training.

$$Loss = \left( -\sum_{i=1}^{Y} \tau_{w_i} \cdot \log(\widehat{\tau_{w_i}}) \right) + \left( -\sum_{j=1}^{Y} \tau_{t_j} \cdot \log(\widehat{\tau_{t_j}}) \right)$$

## Inference

During inference, we send the same data sets used for training to the trained models to generate outputs and make inferences. Recall each model has two outputs. The first output predicts the middle terms, and the second output predicts the execution times of the middle terms. Rather than retrieving these predicted values, we retrieve each model's output arrays from which these predictions are made. The first array shows the probabilities of all terms (in the vocabulary) while the second array shows the probabilities of all times (in hours) at which the term is executed. From these arrays, we retrieve the probabilities of the actual term $P(w)$ and the actual time $P(\tau)$.

Next, we create two matrices for each task. The first one ($\Gamma_{H_i}^{Y_{i,j}}$) contains the probability values of the terms, and the second one ($\Gamma_{\tau}^{Y_{i,j}}$) contains the probability values of the execution times. Then, we compute the mean probabilities of the two matrixes as follows.

$$\rho_{i,j}^{w} = \frac{\sum_{\forall w \in \Gamma_{w}^{Y_{i,j}}} P(w)}{|\Gamma_{w}^{Y_{i,j}}|} \quad , \quad \rho_{i,j}^{\tau} = \frac{\sum_{\forall w \in \Gamma_{\tau}^{Y_{i,j}}} P(\tau)}{|\Gamma_{\tau}^{Y_{i,j}}|}$$

The values $\rho_{i,j}^{w}$ and $\rho_{i,j}^{\tau}$ determine the likelihood that the task $Y_{i,j}$ occurs in the data set and the likelihood that $Y_{i,j}$ was executed at the time $\tau$, respectively. Finally, we sort the mean probabilities $\rho_{i,j}^{w}$ and $\rho_{i,j}^{\tau}$ for $j \in [1, n]$ separately in ascending order for each host $H_i$. To reconcile the differences between these two ordered lists, we identify the minimum ranking of the two. Hence, the higher the ranking, the more likely that a task is an anomaly.

# Experiment

We evaluated efficacy of this framework by setting the inactivity period ($\delta$) to 3 minutes, the number of terms ($\gamma$) to 20, and the window size ($w$) to 5. To ensure the validity of the results, we ran the experiment three times. For each trial, we trained the model for 30 epochs with early stopping, which stopped the training after five iterations if the loss value did not improve by more than 0.0001.

## Data Set

The data used in this study is obtained exclusively from a security operations center that provides services to clients globally. The data was collected over a period of two weeks from a large corporation using a SIEM tool. It includes a total of 904,120 process audits conducted on 605 distinct IP addresses. We assume that each IP address corresponds to a single host.

As outlined in the data preprocessing section, we extracted the command that initiated each process and identified the tasks performed on each host. It is important to reiterate that each task comprises multiple commands. For our analysis, we chose to focus on the tasks of only 118 hosts. The selection of these hosts was based on two factors: 1) a host must have had 100 or more tasks, and 2) a host must have been idle during non-business hours, ensuring that it is an endpoint device of an employee. As a result, we used a total of 33,933 tasks from the 118 hosts, averaging 288 tasks per host.

## Injection Schemes

To test the effectiveness of our framework, we injected anomalous tasks and evaluated how well our framework detected them. We employed three injection schemes, with the first one involving the tasks. For each host $H$, we randomly selected a task from another host and measured its cosine similarity against $H$'s all tasks based on the document-term matrices of the tasks. To make it more challenging and realistic, we ensured that the injected task was slightly different from $H$'s tasks but not entirely unique. This meant that the randomly selected task's similarity with $H$'s tasks was between 1-45 degrees angle, corresponding to a cosine similarity value between 0.71 and 1.00. If the randomly selected task did not satisfy this criterion, we repeated this process until one was identified. We performed this procedure for each of the 118 hosts.

The second injection scheme focused on the execution times of tasks. For each host $H$, we randomly selected a task from $H$'s own tasks and modified its execution time to make it appear as though the task was executed outside $H$'s regular operating hours. Selecting from $H$'s own tasks made the injection scheme more challenging and realistic. As before, we repeated this procedure for each of the 118 hosts.

The third injection scheme combined the first and second injection schemes. For each host $H$, we injected the tasks used in the first injection scheme but modified their execution times to match those used in the second injection scheme. This ensured that the tasks, as well as their execution times, were different from all the host's existing tasks. We repeated this for each of the 118 hosts.

It is important to note that we conducted separate analyses for each of these injection schemes. We report our findings separately in the Results section.

## Evaluation

To assess the effectiveness of our framework, we measure the percentile at which an injection can be detected. We determine this value, for each host, by sorting the tasks based on their probability of being an anomaly, from highest to lowest, and then calculating the percentile at which the injection is detected for each injection scheme. To ensure the validity of our results, we repeat this process three times for each injection scheme. Finally, we report the average percentile at which the injections are identified. Lower percentiles indicate better performance.

## Benchmark Models

The data set used in this study is unique and the problem being investigated is unsupervised in nature. This makes it challenging to identify benchmark models from earlier studies. Previous studies do not focus on examining SIEM logs, except for one study by Goldstein et al. (2013), which does not investigate process audits. Other studies examine different types of text-based logs such as HPC logs (e.g., Baseman et al., 2016; Guo, Yuan, & Wu, 2021; Haque, DeLucia, & Baseman, 2017; Li et al., 2018; Meng et al., 2019; Wittkopp et al., 2021) or HDFS logs (e.g., Du et al., 2017; Guo, Yuan, & Wu, 2021; Meng et al., 2019; Xu et al., 2009). Further, most of these studies use supervised models (e.g., Goldstein et al., 2013; Guo, Yuan, & Wu, 2021; Haque, DeLucia, & Baseman, 2017; Li et al., 2018; Wittkopp et al., 2021). Therefore, we were unable to find any benchmark models from earlier studies that could be used for comparison in this study. To address this, we developed benchmark models using statistical text mining, which is a well-established approach for processing unstructured and text-based data (see Salton & McGill, 1983).

We built several different models using statistical text mining. Because these models relied on statistical text mining, they were trained using different data sets than those used for our framework. For each host, we created two different document-term matrices, one with binary weights and another with term frequency—inverse document frequency (TF-IDF) weights, to capture the representation of the host's tasks. These matrices were further augmented by a one-hot encoded matrix that captured each task's execution time (in hours). Therefore, we had two feature sets for each host. Each feature set captured weights for the terms and the execution times of every task. We trained separate models using the density-based spatial clustering of applications with noise (DBSCAN) and one-class support vector machine (OC-SVM) algorithms using these feature sets. In DBSCAN, we used epsilon values ranging from 0.5 to 10 (in 0.5 increments). In OC-SVM, we used a gamma value of "auto." We evaluated the effectiveness of these models using the same percentile approach discussed earlier.

## Results

Table 1 presents the results of the first injection scheme. As demonstrated in the table, our framework outperformed the benchmark models. Specifically, our framework detected injected anomalies at an average percentile of 6.46 (with a standard deviation of 8.87) when the data sets were sorted from the most likely to be an anomaly to the least likely. The benchmark model that performed the closest to our framework was the OC-SVM, which used the feature set with TF-IDF term weights and identified anomalies at an average percentile of 7.38 (with a standard deviation of 16.20).

| Injection schema 1 | Mean percentile | Minimum percentile | Maximum percentile | Standard deviation |
|---|---|---|---|---|
| DBSCAN – binary weights (eps=2) | 42.65 | 3.90 | 100 | 25.83 |
| DBSCAN – TF-IDF weights (eps=1.5) | 77.25 | 0.18 | 100 | 41.94 |
| OC-SVM – binary weights | 16.13 | **0.16** | 100 | 24.47 |
| OC-SVM – TF-IDF weights | 7.38 | 0.18 | 100 | 16.20 |
| CBoTT | **6.46** | 0.18 | **45.65** | **8.87** |

| Table 1. Results of the Injection Scheme 1 |
| :---: |

Table 2 presents the results of the second injection scheme, which showed that our framework continued to outperform the benchmark models. Specifically, our framework identified injections at an average percentile of 1.82 (with a standard deviation of 4.79) when the data sets were sorted from the most likely to be an anomaly to the least likely. The closest benchmark model was the OC-SVM, which used the feature set with TF-IDF term weights. This model identified injections at an average percentile of 13.52 (with a standard deviation of 13.33).

| Injection schema 2 | Mean percentile | Minimum percentile | Maximum percentile | Standard deviation |
| :--- | :---: | :---: | :---: | :---: |
| DBSCAN – binary weights (eps=2) | 80.92 | 7.86 | 100 | 28.99 |
| DBSCAN – TF-IDF weights (eps=1.0) | 79.55 | 12.01 | 100 | 26.19 |
| OC-SVM – binary weights | 36.01 | 0.75 | 90.59 | 23.53 |
| OC-SVM – TF-IDF weights | 13.52 | 0.20 | 89.39 | 13.33 |
| CBoTT | **1.82** | **0.16** | **40.72** | **4.79** |
| **Table 2. Results of the Injection Scheme 2** | | | | |

Table 3 displays the results of the third injection scheme. As evident from the table, the OC-SVM model which used a feature set with TF-IDF term weights slightly outperformed our framework. The OC-SVM identified injections at an average percentile of 3.25, whereas our framework identified them at an average percentile of 3.54. However, our framework had a smaller standard deviation of 6.35 compared to the 12.12 observed for the OC-SVM. Additionally, our framework was able to detect injections at a maximum percentile of 45.14, while the OC-SVM was able to do so at 94.64.

| Injection schema 3 | Mean percentile | Minimum percentile | Maximum percentile | Standard deviation |
| :--- | :---: | :---: | :---: | :---: |
| DBSCAN – binary weights (eps=2) | 41.43 | 3.90 | 100 | 24.76 |
| DBSCAN – TF-IDF weights (eps=1.0) | 13.82 | **0.16** | 100 | 33.29 |
| OC-SVM – binary weights | 15.74 | **0.16** | 100 | 24.06 |
| OC-SVM – TF-IDF weights | **3.25** | 0.18 | 94.64 | 12.12 |
| CBoTT | 3.54 | 0.18 | **45.14** | **6.35** |
| **Table 3. Results of the Injection Scheme 3** | | | | |

## Discussion

### *Summary of Findings*

In this paper, we present the unsupervised CBoTT framework designed to conduct threat hunting in logs that contain process audits. The framework works by extracting text-based commands from process audits, pre-processing them for analysis, grouping them into tasks for each host, creating a training dataset for each host using the continuous bag-of-words approach, and identifying anomalous tasks based on the commands executed and the execution time for each task. It is important to note that the presence of anomalous tasks does not necessarily indicate malicious activities, but rather provides a starting point for cybersecurity analysts to investigate the logs and determine whether any malicious intent is present.

To assess the effectiveness of the framework, we calculate the percentile at which it detects injected anomalies after it sorts the data by the likelihood of being an anomaly, from most to least likely. Additionally, we compare the framework's performance to several benchmark models built using statistical text mining.

The results of our analyses for three different injection schemes show that the CBoTT framework can identify anomalies at an average percentile range of 1.82 to 6.46. This is an improvement compared to the benchmark models, which can detect anomalies at an average percentile range of 3.25 to 80.92. Overall, the CBoTT framework demonstrates superior performance compared to the benchmark models.

## Limitations

Our work is not without limitations. First, the framework only identifies anomalous tasks without distinguishing between malicious and benign anomalies. This means that a task flagged as anomalous by the framework may be benign, which may cause analysts to investigate non-critical tasks during threat hunting. When there are too many non-critical and benign anomalies, it can lead to a phenomenon called alert fatigue, where analysts become desensitized to the alerts being generated. Therefore, analysts' responsiveness and ability to identify critical and malicious activities can be impaired. However, this limitation is primarily due to the unsupervised nature of the framework. It is possible for the framework to distinguish between malicious and benign tasks if it is trained on labeled data so that it learns the representations of malicious tasks. However, in real-world settings, the nature of malicious activities can change rapidly and vary from one context to another, making it impractical to label data.

Second, the framework assumes that malicious tasks are unique and one-of-a-kind, making them anomalies in the logs. However, if the same malicious activity occurs frequently in the logs, the framework may learn the representation of this activity and treat it as a commonly occurring task. As a result, the percentile at which this task can be identified could be much higher, depending on the number of times it repeats in the logs. To address this limitation, we can sort the results of the framework in reverse order, from least to most likely to be an anomaly. This will highlight the most frequently occurring tasks in the logs, allowing analysts to determine if any of these tasks require further investigation.

## Implications

Data breaches are a growing problem for organizations with costly consequences. It can take up to nine months to identify a breach, and during this time, cyber attackers can steal valuable information resources, causing financial and reputational damage. One method to detect breaches, and other types of cyber-attacks, is threat hunting, which involves analyzing system logs generated by endpoint devices. However, threat hunting can be challenging because of the large size of logs. Cybersecurity analysts might need to analyze hundreds of gigabytes of log files to determine if there are any suspicious activities. If the logs contain process audits, which are text-based commands executed by users or applications on endpoint devices, it becomes even more challenging to identify malicious activities.

The CBoTT framework proposed in this study can effectively address these challenges by analyzing large volumes of text-based commands and identifying unique and suspicious activities. This framework is unsupervised, which means that it does not require analysts to label data. However, it is essential to note that activities identified by CBoTT are not necessarily malicious. Instead, CBoTT identifies activities that do not frequently occur on endpoint devices. Cybersecurity analysts should examine these activities and determine those that pose risks to an organization. They can then invoke threat response procedures to contain these risks and reduce their impact.

The CBoTT framework can be used in settings beyond cybersecurity where analysis of text-based data is necessary but labeled data is limited or unavailable. For instance, in the healthcare industry, patient histories are often recorded in text-based documents. The framework can be used to identify patients with unique complaints so that their care can be reassessed. Similarly, in online forums, the framework can be used to identify infrequently occurring messages and prevent potentially harmful messages from being posted. The framework can also be deployed on endpoint devices such as smartphones to analyze communications on platforms that provide end-to-end encryption. It can identify phishing or other malicious attempts by analyzing a user's regular communication pattern and automatically flagging irregular communications so the user can employ caution. Overall, the CBoTT framework can be easily used in any setting where text-based data must be analyzed to identify unique or irregular patterns.